\documentclass[twocolumn,english,3p,preprint]{elsarticle}
\usepackage[T1]{fontenc}
\usepackage[latin9]{inputenc}
\usepackage{amsmath}
\usepackage{graphicx}
\usepackage{amssymb}
\usepackage{babel}

\begin{document}

\title{Light Scattering on Nanowire Antennas:\\
 A Semi-Analytical Approach}

\author{C. Kremers}

\ead{kremers@uni-wuppertal.de}

\author{D. N. Chigrin}

\ead{chigrin@uni-wuppertal.de}

\address{Institute of High-Frequency and Communication Technology, Faculty
of Electrical, Information and Media Engineering, University of Wuppertal,
Rainer-Gruenter-Str.21, D-42119 Wuppertal, Germany}
\begin{abstract}
Two semi-analytical approaches to solve the problem of light scattering
on nanowire antennas are developed and compared. The derivation is
based on the exact solution of the plane wave scattering problem in
case of an infinite cylinder. The original three-dimensional problem
is reduced in two alternative ways to a simple one-dimensional integral
equation, which can be solved numerically by a method of moments approach.
Scattering cross sections of gold nanowire antennas with different
lengths and aspect ratios are analysed for the optical and near-infrared
spectral range. Comparison of the proposed semi-analytical methods
with the numerically rigorous discrete dipole approximation method demonstrates
good agreement as well as superior numerical performance.\end{abstract}
\maketitle

\section{Introduction}

Since decades antennas are used in everyday devices in the radio and
microwave spectral range as a bridge between propagating radiation
and localized fields. In this spectral range semi-analytical models
exist \citep{Balanis1989a}, offering insight into interaction processes
and guiding engineers in antenna design. In the same time, for practical
antenna design and optimization well established numerical tools are
typically used \citep{Volakis2007}. Shifting antenna resonances towards
optical spectral range brings new challenges both from the fabrication
and the theoretical perspectives. At optical frequencies metal can
no longer be treated as perfect electric conductor and dimensions
of the antenna might be as small as several tens of nanometers \citep{Bharadwaj2009}.
Recent progresses in nanotechnology have enabled the fabrication of
optical antennas (nano-antennas) \citep{Bharadwaj2009}, \citep{Muhlschlegel2005}
and opened many exciting possibilities towards nano-antenna applications.
For example, it has been recently demonstrated that nano-antennas
can enhance \citep{Rogobete2007} and direct the emission of single
molecules \citep{Taminiau2008} and that they can play a key role
in sensing application \citep{Raschke2003}. Great potential in improving
the efficiency of solar-cells should also be mentioned \citep{Atwater2010}.

Design and optimization of optical antennas are mainly done using
general numerical Maxwell solvers \citep{myro2008}, which demand
huge computational resources. Therefore accurate analytical and semi-analytical
models predicting characteristics and performance of nano-antenna
are of great importance. There are just very few exact analytical
solutions available. Light scattering on metal spheres \citep{Bohren1998},
infinite long cylinders \citep{Bohren1998} and spheroids \citep{Sinha1977}
can be derived in closed analytical form. For the light scattering
problem involving a finite length nanowire, a Pocklington-like one-dimensional
(1D) integral equation can be introduced using equivalent surface
impedance method \citep{Hanson2006}. In this paper, we propose a
further development of this approach, which provides better accuracy
especially in the case of nano-antennas with small aspect ratio.

The paper is organized as follows. In section 2 the problem of light
scattering on a thin perfectly conducting wire is reviewed. Pocklington's
integral equation is introduced and extended to the case of a nanowire
of finite conductivity. We demonstrate how one can use the knowledge
of the exact solution of the problem of the plane wave scattering
on an infinite cylinder in order to improve accuracy of the surface
impedance method \citep{Hanson2006}. In Sections 3 and 4 a new numerical
method to solve the resulting one-dimensional (1D) integral equation
is introduced. The method involves a method of moments (MoM) like
discretization scheme, but does not require any specific boundary
conditions to be imposed at the nanowire ends. In section 5 numerical
calculations of scattering cross-sections for plane wave scattering
on gold nanowires with varying geometries is presented and compared
with numerically rigorous discrete dipole approximation (DDA) calculations
\citep{YURKIN2007a}. Section 6 summarizes the paper.

\section{Pocklington like equation}

\begin{figure}
\begin{centering}
\includegraphics[height=0.6\columnwidth]{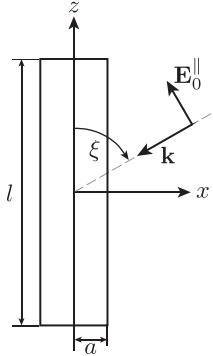}
\par\end{centering}

\caption{\label{fig:geometry}Definition of the geometrical parameters, radius
$a$ and length $l$, of the scattering cylinder as well as the chosen
body centered coordinate system. Additionally the incident angle $\xi$
and the polarization basis vector $\mathbf{E}_{0}^{\parallel}$ are depicted in the incident plane.}
\end{figure}

In infinite free space the electric field $\mathbf{E}(\mathbf{r})$
generated by a time harmonic current density distribution $\mathbf{j}(\mathbf{r})$
(time dependence $e^{-i\omega t}$) enclosed in the finite volume
$V$ is given by \citep{Bladel2007} \begin{multline}
\mathbf{E}(\mathbf{r})=\mathbf{E}^{inc}(\mathbf{r})+i\omega\mu_{0}\left(\overleftrightarrow{\mathbf{I}}+\frac{1}{k^{2}}\nabla\otimes\nabla\right)\\
\int_{V}g(\mathbf{r},\mathbf{r}')\mathbf{j}(\mathbf{r}')\, d^{3}r'\label{eq:integral_equation_j}\end{multline}
 where $\overleftrightarrow{\mathbf{I}}$ denotes the three-dimensional
unit tensor, $\otimes$ is the tensor product defined by $\left(\mathbf{a}\otimes\mathbf{b}\right)_{ij}=a_{i}b_{j}$,
\begin{equation}
g(\mathbf{r},\mathbf{r}')=\frac{e^{ik\left|\mathbf{r}-\mathbf{r}'\right|}}{4\pi\left|\mathbf{r}-\mathbf{r}'\right|}\label{eq:scalar_greens_function}\end{equation}
 is the scalar Green's function and $k=\frac{\omega}{c}$ the free
space wave number. $\mathbf{E}^{inc}$ is an electric field due to
sources not contained in $V$. In scattering problems the driving
current density is not controlled from the outside, but instead induced
by $\mathbf{E}^{inc}$ which plays the role of an incident field in
this case. Spatial derivatives of the integral in (\ref{eq:integral_equation_j})
exist as long as the source current density $\mathbf{j}(\mathbf{r})$
satisfies the H\"{o}lder condition \citep{Kellogg1953}, $\left|\mathbf{j}\left(\mathbf{r}\right)-\mathbf{j}(\mathbf{r}')\right|\leq k\left|\mathbf{r}-\mathbf{r}'\right|^{\alpha}$
with $k>0$ and $0<\alpha\leq1$ for every pair $\mathbf{r},\mathbf{r}'\in V$.

In what follows we consider light scattering on a cylinder (wire)
with length $l$ and radius $a$ (Fig.~\ref{fig:geometry}). The
geometry of the problem including a body centered coordinate system
with $z$-axis parallel to the cylinder axis is depicted in figure~\ref{fig:geometry}.
The wave vector $\mathbf{k}$ enclosing the incident angle $\xi$
with the positive $z$-axis lies in the $xz$-plane (incident plane).
To excite longitudinal resonances only the projection of the incident
plane wave electric field on the incident plane $\mathbf{E}_{0}^{\parallel}=E_{0}^{\parallel}\left(\sin\xi\hat{\mathbf{z}}-\cos\xi\hat{\mathbf{x}}\right)$
have to be taken into account. The perpendicular polarization component can be
safely ignored. Under the assumption that the cylinder diameter is
much smaller than the free space wavelength, i.e. $ka\ll1$, the incident electric field
interacting with the cylinder can be viewed as a function depending
only on $z$\begin{eqnarray}
\mathbf{E}^{inc}(z) & \approx & \mathbf{E}_{0}^{\parallel}e^{-ikz\cos\xi}.\label{eq:incident_field-1}\end{eqnarray}

First we briefly review the derivation of Pocklington's equation for
scattering on a thin perfectly conducting wire \citep{Balanis1989a}.
In this case in cylindrical coordinates $\left\{ \rho,\phi,z\right\} $
the induced current density $\mathbf{j}(\mathbf{r})$ has only a $z$-component,
shows no $\phi$-dependence and exists solely on the antenna interface.
Then the induced current $I(z)$ is related to the current density
$\mathbf{j}(\mathbf{r})$ as\begin{equation}
\mathbf{j}(\mathbf{r})=\hat{\mathbf{z}}I(z)\frac{\delta(\rho-a)}{2\pi a}\label{eq:surface_current_density}\end{equation}
 so that \begin{equation}
I(z)=\int_{0}^{2\pi}d\phi\int_{0}^{a}\rho d\rho\, j(\mathbf{r}).\label{eq:I(z)}\end{equation}
 Using (\ref{eq:surface_current_density}) in (\ref{eq:integral_equation_j})
and performing the $\rho'$-integration one yields for the $z$-component
of the electric field on the cylinder surface the following integro-differential
equation\begin{multline}
E_{z}(a,z)=E_{z}^{inc}(z)+i\frac{\omega\mu_{0}}{2\pi}\left(1+\frac{1}{k^{2}}\frac{\partial^{2}}{\partial z^{2}}\right)\\
\int_{-\frac{l}{2}}^{\frac{l}{2}}dz'\int_{0}^{2\pi}d\phi'\, g_{a}(\phi',z-z')I(z').\label{eq:E_z(rho,z)}\end{multline}
where $g_{a}(\phi',z-z')=g(\mathbf{r},\mathbf{r}')$ with $\mathbf{r}=\left( a,0,z\right) $
and $\mathbf{r}'=\left( a,\phi',z'\right) $ in cylindrical coordinates.
While the antenna is perfectly conducting, the $z$-component of the
total field at the antenna surface has to vanish, $E_{z}(a,z)=0$.
Applying this boundary condition to Eq.~(\ref{eq:E_z(rho,z)}) one
can derive Pocklington's integro-differential equation in the form\begin{multline}
E_{z}^{inc}(z)=-i\frac{\omega\mu_{0}}{2\pi}\left(1+\frac{1}{k^{2}}\frac{\partial^{2}}{\partial z^{2}}\right)\\
\int_{-\frac{l}{2}}^{\frac{l}{2}}dz'\int_{0}^{2\pi}d\phi'\, g_{a}(\phi',z-z')I(z'),\label{eq:IE_pec}\end{multline}
 which provides a general solution of the scattering problem \citep{Balanis1989a}.
Here the fact, that the antenna is perfectly conducting, is taken
into account twice, first by assuming a special form of the induced
current (\ref{eq:surface_current_density}) and second by enforcing
the boundary condition $E_{z}(a,z)=0$.

A typical nano-antenna at optical frequency range possesses high,
but finite conductivity. In this case the surface current approximation
(\ref{eq:surface_current_density}) is still applicable, while the
boundary condition $E_{z}(a,z)=0$ is generally not. In order to use
a Pocklington's like equation at this frequency range, one needs a
relationship between the field $E_{z}(a,z)$ and the current $I(z)$
at the antenna interface.

Assuming a long nanowire, i.e. $l\gg a$, it is reasonable to expect
that the internal electric field is separable similar to the solution
of the equivalent problem involving an infinite cylinder \citep{Bohren1998}.
Moreover, it can be shown that as long as the wire is electrically
thin, i.e. $ka\ll1$, and $\left|\epsilon_{r}\right|\gg1$ at least for the case of inclined incidents
($\xi\neq\frac{\pi}{2}$) the $\rho$- and $\phi$-components of the
internal field are negligible compared to the $z$-component of the
field and the $z$-component shows no $\phi$-dependence. Assuming
these conditions are fulfilled the internal electric field can be
approximated by \begin{equation}
\mathbf{E}(\mathbf{r})\approx\hat{\mathbf{z}}f(z)J_{0}\left(k_{\rho}\rho\right)\label{eq:field_ansatz}\end{equation}
 where $J_{0}\left(k_{\rho}\rho\right)$ denotes the Bessel function
of the first kind, $k_{\rho}=k\sqrt{\epsilon_{r}-\cos^{2}\xi}$ with
relative permittivity $\epsilon_{r}$ of the wire and $f(z)$
an unknown $z$-dependent function giving the amplitude of the internal
field along the wire. A connection between the induced current density
and the internal electric field is given by means of the volume equivalence
theorem \citep{Balanis1989a} by\begin{equation}
\mathbf{j}(\mathbf{r})=-i\omega\epsilon_{0}\Delta\epsilon_{r}\mathbf{E}(\mathbf{r})\label{eq:surface equivalence theorem}\end{equation}
 with $\Delta\epsilon_{r}=\epsilon_{r}-1$. Combining (\ref{eq:field_ansatz})
and (\ref{eq:surface equivalence theorem}) the total current through
the wire can be calculated \begin{align}
I(z) & =\int_{0}^{2\pi}d\phi\int_{0}^{a}\rho d\rho\, j_{z}(\rho,z)\nonumber \\
 & =-i\omega\epsilon_{0}\Delta\epsilon_{r}2\pi a\frac{J_{1}\left(k_{\rho}a\right)}{k_{\rho}}f(z).\label{eq:I(z) infinite cylinder}\end{align}
 Further comparing results of the integration in (\ref{eq:I(z) infinite cylinder})
with (\ref{eq:field_ansatz}) one can derive the following relation
between the electric field and the total current at the wire interface
\begin{equation}
E_{z}(a,z)=Z_{S}I(z)\label{eq:E_from_impedance}\end{equation}
 with the surface impedance\begin{equation}
Z_{S}=i\frac{J_{0}(k_{\rho}a)k_{\rho}}{2\pi a\omega\epsilon_{0}\Delta\epsilon_{r}J_{1}(k_{\rho}a)}.\label{eq:surface impedance}\end{equation}
 Using relation (\ref{eq:E_from_impedance}) the following Pocklington
like integro-differential equation for induced total current, the
surface impedance (SI) integro-differential equation \citep{Hanson2006},
can be obtained from equation (\ref{eq:E_z(rho,z)}) \begin{multline}
Z_{S}I(z)=E_{z}^{inc}(z)+i\frac{\omega\mu_{0}}{2\pi}\left(1+\frac{1}{k^{2}}\frac{\partial^{2}}{\partial z^{2}}\right)\\
\int_{-\frac{l}{2}}^{\frac{l}{2}}dz'\int_{0}^{2\pi}d\phi'\, g_{a}(\phi',z-z')I(z').\label{eq:SI-IE}\end{multline}

We propose a further improvement to the approximation (\ref{eq:SI-IE})
by releasing the solely surface current ansatz (\ref{eq:surface_current_density}).
In order to do that, we assume that the induced current density on
the right hand side of equation (\ref{eq:integral_equation_j}) can
be factorize similar to the internal field (\ref{eq:field_ansatz}).
In this way substituting (\ref{eq:field_ansatz}) in equation (\ref{eq:integral_equation_j})
both on the left hand side as boundary condition as well as by using
(\ref{eq:surface equivalence theorem}) on the right hand side to
rewrite the induced current density one obtains a self-consistent
integro-differential equation for the unknown amplitude $f(z)$ \begin{multline}
f(z)J_{0}\left(k_{\rho}a\right)=E_{z}^{inc}(z)+k^{2}\Delta\epsilon_{r}\left(1+\frac{1}{k^{2}}\frac{\partial^{2}}{\partial z^{2}}\right)\\
\int_{V}d^{3}r'\, g(a,z;\mathbf{r}')f(z')J_{0}\left(k_{\rho}\rho'\right)\label{eq:VC-IE}\end{multline}
where $g(a,z;\mathbf{r}')=g(\mathbf{r},\mathbf{r}')$ with $\mathbf{r}=\left( a,0,z\right) $
in cylindrical coordinates. This volume current (VC) integro-differential
equation takes into account both appropriate boundary conditions at
the wire interface and an appropriate volume current distribution
inside the wire.

\section{Discrete form of VC integral equation\label{sec:Solving-the-Volume}}

In order to solve numerically the integro-differential equations (\ref{eq:SI-IE})
and (\ref{eq:VC-IE}) one has to impose additional boundary conditions
at the nano-antenna edges. A common choice is to impose the total
current $I(z)$ to be equal to zero for $z=\pm l/2$ \citep{Hanson2006}.
For a solid wire with finite conductivity this choice is generally
not justified, while the total current can be discontinuous at the
wire edges \citep{Bladel2007}.

To overcome the requirement of additional boundary condition one has
to convert the integro-differential equations into purely integral
ones. To do that one has to bring the differential operator $\left(1+k^{-2}\partial_{z}^{2}\right)$
inside the integral. This procedure results in singularities of the
order $\left|\mathbf{r}-\mathbf{r}'\right|^{3}$ which are generally
not integrable over a volume. To treat these singularities we follow
the regularization scheme proposed by Lee et al. in Ref.~\citep{Lee1980}.
The regularized VC equation (\ref{eq:VC-IE}) reads: \begin{multline}
\left(1+\Delta\epsilon_{r}L_{33}\right)E_{z}(a,z)=E_{z}^{inc}(z)+\\
k^{2}\Delta\epsilon_{r}\left\{ \int_{V-V^{\star}}G_{33}(a,z;\mathbf{r}')E_{z}(\rho',z')d^{3}r'+\right.\\
\int_{V^{\star}}\Biggl[G_{33}(a,z;\mathbf{r}')E_{z}(\rho',z')-\\
\frac{1}{k^{2}}\frac{\partial^{2}}{\partial z^{2}}g_{0}(a,z;\mathbf{r}')E_{z}(a,z)\Biggr]\, d^{3}r'\Biggr\}\label{eq:volume_integral_equation}\end{multline}
 with $E_{z}(\rho,z)=f(z)J_{0}\left(k_{\rho}\rho\right)$. Here $V^{\star}$
denotes a finite and arbitrary shaped principal volume $V^{\star}$
containing the singular point $\mathbf{r}=(a,0,z)$, $g_{0}(\mathbf{r},\mathbf{r}')=\lim_{k\rightarrow0}g(\mathbf{r},\mathbf{r}')$
is the static scalar Green's function, $G_{33}$ is the $zz$-element
of the dyadic Green's function \citep{Bladel2007} \begin{equation}
G_{33}(\mathbf{r},\mathbf{r}')=\left(1+\frac{1}{k^{2}}\frac{\partial^{2}}{\partial z^{2}}\right)g(\mathbf{r},\mathbf{r}'),\label{eq:dyadic_greens_function}\end{equation}
 and $L_{33}$ the $zz$-element of the source dyadic \citep{Bladel2007}
\begin{equation}
L_{33}=\frac{1}{4\pi}\oint_{\partial V^{\star}}d^{2}r'\,\frac{\left(z'-z\right)\left(\hat{\mathbf{n}}\cdot\hat{\mathbf{z}}\right)}{\left|\mathbf{r}-\mathbf{r}'\right|^{3}}.\label{eq:source_dyadic}\end{equation}
 The surface integration in (\ref{eq:source_dyadic}) has to be performed
over the surface $\partial V^{\star}$ enclosing the principal volume
$V^{\star}$, $\hat{\mathbf{n}}$ denotes the outer surface normal.

\begin{figure}
\begin{centering}
\includegraphics[width=0.8\columnwidth]{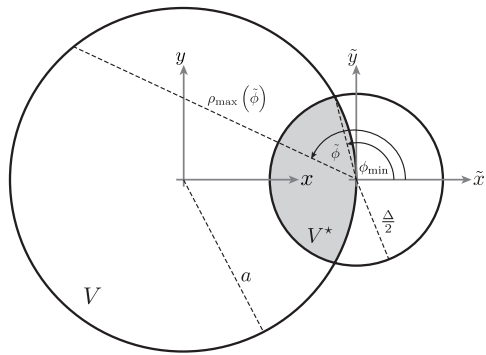}
\par\end{centering}

\caption{\label{fig:coordinate-trafo}Cross-section of the nanowire (big circle)
and the principal volume $V^{\star}$ (small circle) together with
the definition of the coordinate systems used to calculate integrals
in equation (\ref{eq:volume_integral_1d-2}).}
\end{figure}

The main advantage of the regularization scheme \citep{Lee1980} is
(i) that all singularities disappear and (ii) that the principal volume
$V^{\star}$ can be finite and arbitrary shaped. We choose a cylinder
with length $\Delta$, radius $\frac{\Delta}{2}$ and center at the
singular point $\mathbf{r}=(a,0,z)$ as the principal volume (Fig.~\ref{fig:coordinate-trafo}).
We assume $\Delta$ to be small, such that both the Bessel function
$J_{0}\left(k_{\rho}\rho'\right)$ and the amplitude function $f(z')$
are approximately constant over $V^{\star}$. Taking that into account,
writing the volume integrals in the regularized VC equation (\ref{eq:volume_integral_equation})
in cylinder coordinates $\left\{ \rho,\phi,z\right\} $ and collecting
all terms containing $f(z)$ on the left and all terms containing $f(z'\neq
z)$ on the right hand side, one obtains a one-dimensional
integral equation in the form \begin{multline}
f(z)\Gamma\approx E_{z}^{inc}(z)+\int_{-\frac{l}{2}}^{z-\frac{\Delta}{2}}dz'f(z')\mathcal{L}(\left|z-z'\right|)+\\
\int_{z+\frac{\Delta}{2}}^{\frac{l}{2}}dz'f(z')\mathcal{L}(\left|z-z'\right|),\label{eq:volume_integral_1d-2}\end{multline}
 with \begin{equation}
\Gamma=J_{0}\left(k_{\rho}a\right)\left\{ 1+\Delta\epsilon_{r}L_{33}-\eta_{in}\right\} -\eta_{out},\label{eq:def_Gamma}\end{equation}
 \begin{multline}
\eta_{in}=4k^{2}\Delta\epsilon_{r}\int_{0}^{\frac{\Delta}{2}}d\tilde{z}\\
\left\{ \left(\int_{\phi_{min}}^{\pi}d\tilde{\phi}\int_{0}^{\frac{\Delta}{2}}\tilde{\rho}d\tilde{\rho}+\int_{\frac{\pi}{2}}^{\phi_{min}}d\tilde{\phi}\int_{0}^{\rho_{max}\left(\tilde{\phi}\right)}\tilde{\rho}d\tilde{\rho}\right)\right.\\
\left.\left(G_{33}(\mathbf{0},\tilde{\mathbf{r}})-\frac{1}{k^{2}}\frac{\partial^{2}}{\partial z^{2}}g_{0}(\mathbf{0},\tilde{\mathbf{r}})\right)\right\} ,\label{eq:def_eta_in}\end{multline}
 \begin{multline}
\eta_{out}=4k^{2}\Delta\epsilon_{r}\int_{0}^{\frac{\Delta}{2}}d\tilde{z}\int_{\phi_{min}}^{\pi}d\tilde{\phi}\int_{\frac{\Delta}{2}}^{\rho_{max}(\tilde{\phi})}\tilde{\rho}d\tilde{\rho}\\
\left\{ J_{0}\left[k_{\rho}\rho'(\tilde{\rho},\tilde{\phi})\right]G_{33}(\mathbf{0},\tilde{\mathbf{r}})\right\} \label{eq:def_eta_out}\end{multline}
 and

\begin{multline}
\mathcal{L}(\left|z-z'\right|)=2k^{2}\Delta\epsilon_{r}\int_{0}^{a}\rho'd\rho'J_{0}\left(k_{\rho}\rho'\right)\\
\int_{0}^{\pi}d\phi'G_{33}(a,z;\rho',z').\label{eq:def_L}\end{multline}
 In (\ref{eq:def_eta_in}) the integration is performed over the principal
volume $V^{\star}$, while in (\ref{eq:def_eta_out}) over the corresponding
wire slice of thickness $\Delta$ centered at $z=z'$ but with excluded
principal volume $V^{\star}$. To calculate these two integrals a
new coordinate system with its center at the singular point has been
chosen Fig.~(\ref{fig:coordinate-trafo}). In the new coordinate
system the source dyadic (\ref{eq:source_dyadic}) can be explicitly
written as\begin{multline}
L_{33}=\frac{(\pi-\phi_{min})\left(2-\sqrt{2}\right)}{2\pi}+\frac{\left(\phi_{min}-\frac{\pi}{2}\right)}{\pi}-\\
\frac{\Delta}{\pi}\int_{\frac{\pi}{2}}^{\phi_{min}}d\tilde{\phi}\frac{1}{\sqrt{4\rho_{max}^{2}(\tilde{\phi})+\Delta^{2}}}.\label{eq:L33}\end{multline}
 The radius vector $\rho'$ and the integration ranges in (\ref{eq:def_eta_in},\ref{eq:def_eta_out},\ref{eq:L33})
parametrically depend on the nanowire radius $a$ and the wire slice
thickness $\Delta$ and are given by\begin{equation}
\rho'(\tilde{\rho},\tilde{\phi})=\left|\left(\begin{array}{c}
a\\
0\end{array}\right)+\left(\begin{array}{c}
\tilde{\rho}\cos\tilde{\phi}\\
\tilde{\rho}\sin\tilde{\phi}\end{array}\right)\right|,\label{eq:rho_transformed}\end{equation}

\begin{align}
\rho_{max}(\tilde{\phi}) & =-2a\cos\tilde{\phi}\nonumber \\
\phi_{min} & =\arccos\left[-\frac{\Delta}{4a}\right].\label{eq:int_range_functions}\end{align}

Taking into account that the nanowire diameter is small in comparison
to the wavelength one can further simplify the integrals in (\ref{eq:def_eta_in})
and (\ref{eq:def_eta_out}) by expanding the Green's function $G_{33}$
in power of $k$ up to the linear term\begin{equation}
G_{33}^{NF}(R)\approx\frac{1}{4\pi}\left(\frac{3R_{z}^{2}-R^{2}}{k^{2}R^{5}}+\frac{R_{z}^{2}+R^{2}}{2R^{3}}+i\frac{2}{3}k\right)\label{eq:G33_NF}\end{equation}
 where $\mathbf{R}=\mathbf{r}'-\mathbf{r}$. Using (\ref{eq:G33_NF})
one can analytically integrate equation (\ref{eq:def_eta_in}) over
$\tilde{z}$ and $\tilde{\rho}$ and equation (\ref{eq:def_eta_out})
over $\tilde{z}$ . Residual integration in (\ref{eq:def_eta_in}),
(\ref{eq:def_eta_out}) and (\ref{eq:L33}) has to be performed numerically.
The regularization scheme (\ref{eq:volume_integral_equation}) ensures,
that numerical integration converges. In this way $\Gamma$ in (\ref{eq:volume_integral_1d-2})
can be efficiently calculated once for given radius $a$ of the nanowire
antenna, $\Delta$ and wavelength.

Equation (\ref{eq:volume_integral_1d-2}) can be solved numerically
by a method of moments (MoM) approach. Specifically we choose a point
matching MoM scheme with pulse-function basis \citep{Orfanidis2010}.
That means we divide the nanowire into a set of slices with thickness
$\Delta$ and label each slice with an index $i$. For reasonable
thin slices the amplitude function $f(z)$ on each of them can be
approximated by its value at the center $f(z_{i})=\mathbf{f}_{i}$.
This leads to the $n=\frac{l}{\Delta}$ dimensional matrix equation
as discretized version of (\ref{eq:volume_integral_1d-2})\begin{equation}
\left(\Gamma\overleftrightarrow{\mathbf{I}}-\overleftrightarrow{\mathbf{M}}\right)\mathbf{f}=\mathbf{E}^{inc}\label{eq:1d_MoM}\end{equation}
 where $\mathbf{E}_{i}^{inc}=E_{z}^{inc}(z_{i})$ and \begin{equation}
M_{ij}=\begin{cases}
\int_{z_{j}-\frac{\Delta}{2}}^{z_{j}+\frac{\Delta}{2}}dz'\mathcal{L}\left(\left|z_{i}-z'\right|\right) & \text{for }i\neq j\\
0 & \text{for }i=j\end{cases}\label{eq:M_ij}\end{equation}
 which can be easily inverted numerically to yield the discrete set
of values $f(z_{i})$. While the Green's function depends on the distance
between two slices $\left|z-z'\right|$, only the first row $M_{1j}$
of the matrix $M$ has to be calculated. All other elements can be
filled using the rule $M_{ij}=M_{i-1,j-1}$.

Using the Green's function expansion (\ref{eq:G33_NF}) in the calculation
of the matrix elements $M_{1j}$ for $j<\frac{2a}{\Delta}$ the integration
over $z'$ can be done analytically. Only the integration over the
nanowire profile has to be done numerically. In the calculation of
the matrix elements $M_{1j}$ for $j>\frac{2a}{\Delta}$ the full
dyadic Green's function have to be used and so the complete volume
integral has to be done numerically. These integrations are done over
$\mathbf{r}'$-regions far from the singularity and they demonstrate
good convergence. An additional performance improvement can be achieved
by expanding the Bessel function $J_{0}$ for small arguments as \citep{Stegun1965}
\begin{equation}
J_{0}(x)\approx1-\frac{x^{2}}{4}+\frac{x^{4}}{64}+\mathcal{O}(x^{6}).\label{eq:bessel_expansion}\end{equation}
 Having the amplitude function $f(z)$ calculated from (\ref{eq:1d_MoM})
the $z$-component of the induced electric field is given by the ansatz
(\ref{eq:field_ansatz}).

\section{Discrete form of SI integral equation}

To solve the surface impedance integro-differential equation (\ref{eq:SI-IE})
the regularization and discretization scheme presented in section~\ref{sec:Solving-the-Volume}
can be also applied. A principal volume is chosen as a full slice
of the wire at position $z$ with thickness $\Delta$. Regularizing
the SI equation (\ref{eq:SI-IE}), writing it in cylinder coordinates
and collecting all terms containing $I(z)$ and $I(z'\neq z)$, one
obtains a one-dimensional integral equation in the form\begin{multline}
I(z)\Gamma^{S}\approx E_{z}^{inc}+\int_{-\frac{l}{2}}^{z-\frac{\Delta}{2}}dz'I(z')\mathcal{L}^{S}(\left|z-z'\right|)+\\
\int_{z+\frac{\Delta}{2}}^{\frac{l}{2}}dz'I(z')\mathcal{L}^{S}(\left|z-z'\right|)\label{eq:Iz_integral_equation_2nd}\end{multline}
 with\begin{equation}
\mathcal{L}^{S}(\left|z-z'\right|)=i\frac{\omega\mu_{0}}{\pi}\int_{0}^{\pi}d\phi'\, G_{33}^{a}(\phi',z-z')\label{eq:LLS}\end{equation}
 and\begin{multline}
\Gamma^{S}=Z_{S}+i\frac{\omega\mu_{0}}{2\pi}\left\{ \frac{1}{k^{2}}L_{33}^{S}-4\int_{0}^{\frac{\Delta}{2}}dz'\int_{0}^{\pi}d\phi'\right.\\
\left.\left[G_{33}^{a}(\phi',z')-\frac{1}{k^{2}}\frac{\partial^{2}}{\partial z^{2}}g_{0}^{a}(\phi',z')\right]\right\} ,\label{eq:GammaS}\end{multline}
 where\begin{equation}
L_{33}^{S}=\frac{\Delta}{4\pi}\int_{0}^{2\pi}d\phi'\,\frac{1}{\sqrt{\left(\frac{\Delta}{2}\right)^{2}+4a^{2}\sin^{2}\left(\frac{\phi'}{2}\right)}^{3}}\label{eq:L33_1st}\end{equation}
and $G_{33}^{a}(\phi',z-z')=G_{33}(\mathbf{r},\mathbf{r}')$ as well
as $g_{0}^{a}(\phi',z-z')=\lim_{k\rightarrow0}g(\mathbf{r},\mathbf{r}')$
with $\mathbf{r}=\left\{ a,0,z\right\} $ and $\mathbf{r}'=\left\{ a,\phi',z'\right\} $
in cylinder coordinates. The source dyadic $L_{33}^{S}$ can be expressed
in terms of the complete elliptic integral of the second kind $E\left(m\right)$
\citep{Stegun1965} as \begin{equation}
L_{33}^{S}=\frac{1}{\pi}\frac{8}{16a^{2}+\Delta^{2}}E\left(-\frac{16a^{2}}{\Delta^{2}}\right),\label{eq:L33_2nd}\end{equation}
 where $E\left(m\right)$ is defined by\begin{equation}
E\left(m\right)=\int_{0}^{\frac{\pi}{2}}\sqrt{1-m\sin^{2}\theta}\, d\theta.\label{eq:Def_elliptic_integral}\end{equation}
 In $\Gamma^{S}$ (\ref{eq:GammaS}) the $z'$-integrations can be
performed analytically with the help of the near-field expansion of
the Green's function (\ref{eq:G33_NF}). The residual $\phi'$-integration
has to be done numerically once for given nanowire radius $a$, $\Delta$
and wavelength.

To solve (\ref{eq:Iz_integral_equation_2nd}) numerically we use a
point matching MoM scheme with pulse function basis \citep{Orfanidis2010}
as in section \ref{sec:Solving-the-Volume}. Again the nanowire should
be divided into a set of slices of thickness $\Delta$ yielding the
matrix equation\begin{equation}
\left(\Gamma^{S}\overleftrightarrow{\mathbf{I}}-\overleftrightarrow{\mathbf{M}}^{S}\right)\mathbf{I}=\mathbf{E}^{inc}\label{eq:SI_IE}\end{equation}
 where\begin{equation}
M_{ij}^{S}=\begin{cases}
\int_{z_{j}-\frac{\Delta}{2}}^{z_{j}+\frac{\Delta}{2}}dz'\mathcal{L}^{S}\left(\left|z_{i}-z'\right|\right) & \text{for }i\neq j\\
0 & \text{for }i=j\end{cases}\label{eq:MS_ij}\end{equation}
 and $I_{i}=I(z_{i})$. Similar to the case of volume current integral
equation considered in section \ref{sec:Solving-the-Volume} only
the first row of the matrix $\overleftrightarrow{\mathbf{M}}^{S}$
have to be calculated due to the symmetry of the Green's function.
Further for slices with $\left|z_{1}-z_{j}\right|\leq2a$ the near
field approximation of the Green's function (\ref{eq:G33_NF}) can
be used to perform the $z'$-integrations in (\ref{eq:MS_ij}) analytically.
The remaining integrations as well as the matrix inversion have to
be performed numerically. Good convergence of the integrals is ensured,
while the integrand is evaluated far from the singular point. Numerical
inversion of (\ref{eq:SI_IE}) results in the current $I_{i}=I(z_{i})$
at the discrete number of points $z_{i}$ along the nanowire. Finally
the $z$-component of the induced electric field is given via (\ref{eq:E_from_impedance}).

\section{Numerical results and discussion}

\begin{figure}
\begin{centering}
\includegraphics[width=0.9\columnwidth]{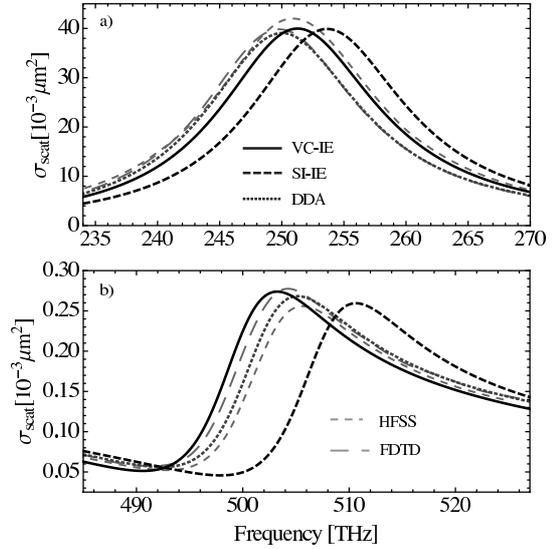}
\par\end{centering}

\caption{\label{fig:comparison_numeric}Scattering cross-section of a gold
nanowire ($l=200nm,\, a=10nm$) under slanting incidence ($\xi=\frac{\pi}{4}$)
calculated with different rigorous numerical methods as well as the
proposed volume current and surface impedance one-dimensional integral
equations. In the top panel (a) the first and in the bottom panel
(b) the third resonance peak are shown.}
\end{figure}

In this section the semi-analytical methods developed in sections
3 and 4 are evaluated and compared with numerically rigorous methods.
Plane wave scattering on a gold nanowire is considered in the optical
and near-infrared spectral range. The relative permittivity of gold
in this spectral range can be modeled by a free electron Drude model\begin{equation}
\epsilon_{r}(\omega)\approx\epsilon_{\infty}-\frac{\omega_{p}^{2}}{\omega^{2}+i\gamma\omega}\label{eq:drude_modell}\end{equation}
 with parameters $\epsilon_{\infty}=9$, $\omega_{p}=1.36674\cdot10^{16}\, s^{-1}$
and $\gamma=7.59297\cdot10^{13}\, s^{-1}$\citep{myro2008}. For comparison
purposes we calculate the total scattering cross-section\begin{equation}
\sigma_{scat}=\frac{\oint_{S}d^{2}r\,\left|\mathbf{E}_{s}\right|^{2}}{\left|\mathbf{E}^{inc}\right|^{2}}\label{eq:sigma_scat_Def}\end{equation}
 where the scattered electric field $\mathbf{E}_{s}$ is generated
by the induced current density $\mathbf{j}(\mathbf{r})$ given by
equation (\ref{eq:surface equivalence theorem}) in the case of the
VC and by equation (\ref{eq:surface_current_density}) in the case
of the SI integral equation model, respectively. In the far-field
zone the scattering field is purely transverse and one can use the
far-field dyadic Green's function\begin{equation}
\lim_{r\rightarrow\infty}\overleftrightarrow{\mathbf{G}}(\mathbf{r},\mathbf{r}')=\frac{e^{ik\left(r-\hat{\mathbf{r}}\cdot\mathbf{r}'\right)}}{4\pi r}\left\{ \overleftrightarrow{\mathbf{I}}-\hat{\mathbf{r}}\otimes\hat{\mathbf{r}}\right\} \label{eq:far_field_greens_function}\end{equation}
 and integral relation\begin{equation}
\mathbf{E}_{s}(\mathbf{r})=i\omega\mu_{0}\int_{V}d^{3}r'\,\overleftrightarrow{\mathbf{G}}(\mathbf{r},\mathbf{r}')\mathbf{j}(\mathbf{r}')\label{eq:scattering_field}\end{equation}
 to calculate the scattering cross-section. For the VC integral equation
model the scattering cross section can be written in terms of the
amplitude function $f(z)$\begin{multline}
\sigma_{scat}=\frac{k^{4}a^{4}\pi}{8\left|\mathbf{E}^{inc}\right|^{2}}\left|\Delta\epsilon_{r}\left(1-\frac{1}{8}a^{2}k^{2}\epsilon_{r}\right)\right|^{2}\\
\int_{0}^{\pi}d\theta\,\sin^{3}\theta\left|\int_{-\frac{l}{2}}^{\frac{l}{2}}dz'f(z')e^{-ik\cos\theta z'}\right|^{2}\label{eq:sigma_scat_VolMoM}\end{multline}
 while for SI integral equation model in terms of the total current
$I(z)$\begin{multline}
\sigma_{scat}^{S}=\frac{\omega^{2}\mu_{0}^{2}}{8\pi\left|\mathbf{E}^{inc}\right|^{2}}\\
\int_{0}^{\pi}d\theta\,\sin^{3}\theta\left|\int_{-\frac{l}{2}}^{\frac{l}{2}}dz'I(z')e^{-ik\cos\theta z'}\right|^{2}.\label{eq:sigma_scat_SI-IE}\end{multline}

\begin{figure}
\begin{centering}
\includegraphics[width=0.9\columnwidth]{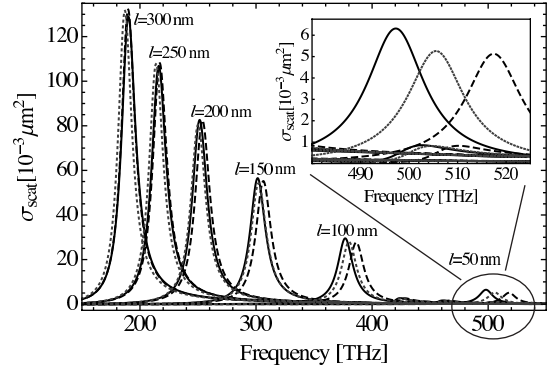}
\par\end{centering}

\caption{\label{fig:1st_resonances}Scattering cross-sections of gold nanowire
with fixed radius $a=10nm$ and different lengths. The full and dashed
lines show the VC and SI integral equation results, respectively.
The dotted line is the numerically rigorous reference calculated using
DDA method.}
\end{figure}

\begin{figure}
\begin{centering}
\includegraphics[width=0.9\columnwidth]{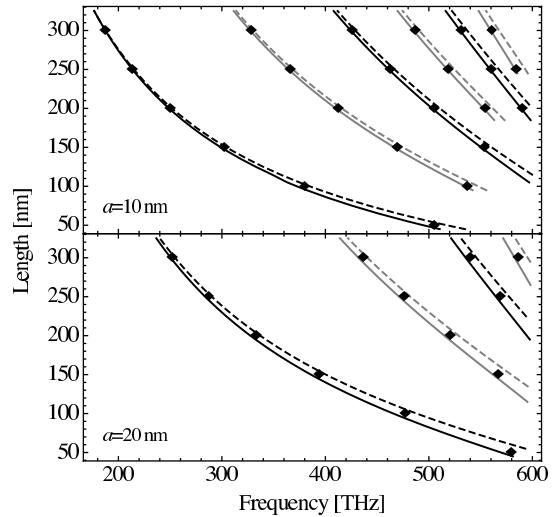}
\par\end{centering}

\caption{\label{fig:map}Contours plot of the resonance frequencies of the
nanowires with different lengths under slanting incidence ($\xi=\frac{\pi}{4}$).
Top panel radius is $a=10nm$, bottom panel $a=20nm$. The full and
dashed lines show the VC and SI integral equation results, respectively.
The dots are represents the numerically rigorous DDA calculations.}
\end{figure}

We have performed a systematic convergence test of the developed one-dimensional
integral equation based methods for gold nanowires of different lengths
and radii. Spatial resolution of $\Delta=1nm$ (thickness of the individual
slices in the wire discretization) typically results in less than
1THz discrepancy with respect to the converged value of the scattering
cross-section maximum calculated at resolution as small as $\Delta=0.05nm$.
In what follows spatial resolution of $\Delta=1nm$ has been used
for all calculations based of the VC and SI integral equation models.

In figure \ref{fig:comparison_numeric} the first (top panel) and
the third (bottom panel) resonance peaks of the cross-section spectra
are shown for a gold nanowire ($a=10nm,\, l=200nm$) under slanting
incidence ($\xi=\frac{\pi}{4}$) with $\left|\mathbf{E}^{inc}\right|=1$.
Results of both different rigorous numerical methods and semi-analytical
1D ones are compared. For the rigorous numerical calculations we used
(i) HFSS, a commercial finite-element frequency-domain Maxwell solver
from ANSYS \citep{HFSS}, (ii) an in-house implementation of the finite
difference time domain method (FDTD) \citep{Taflove2000} and (iii)
ADDA, an open-source software package for calculating scattering parameters
using the discrete dipole approximation (DDA) algorithm \citep{Yurkin2007}.
The space discretization in the shown DDA and FDTD calculations was
set to $1nm$. Discrepancies among the scattering cross-section spectra
calculated using different rigorous three-dimensional (3D) Maxwell
solvers are comparable with discrepancies of these spectra with the
spectrum calculated using VC integral equation method. The accuracy
of the SI integral equation method is slightly worse but still very
reasonable.

Most important are the differences in used computational resources
and execution time. The calculation of one frequency point in VC (SI)
integral equation method requires approximately 2 (1) seconds on one
core of a workstation using Mathematica \citep{mathematica}. In contrast
DDA calculations requires around 8 minutes per frequency point on
the same workstation. HFSS needs around 6 minutes per frequency point
if the mesh is optimized for one frequency only and is reused without
optimization for 30 other frequencies. With FDTD one gets the complete
spectra in one run in around 250 minutes on one core. In conclusion
the newly proposed 1D semi-analytical methods provide a speed-up in
execution time close to 200 times compared to general 3D Maxwell solvers.
Additionally up to 100 times less RAM is required for the semi-analytical
calculations.

In figure~\ref{fig:1st_resonances} the position of the scattering
cross-section maxima are shown for gold nanowires with different aspect
ratios. Radius is $a=10nm$ and length varies from $l=50nm$ (shown
in the inset) to $l=300nm$. Normal incidence ($\xi=\frac{\pi}{2}$)
with $\left|\mathbf{E}^{inc}\right|=1$ is considered. VC integral equation method (full black line), SI integral
equation method (dashed line) and DDA (dotted line) are compared.
As it is expected the accuracy of both semi-analytical methods deteriorates
with decreasing aspect ratio. However even for considerable low aspect
ratio $\frac{5}{4}$ ($l=50nm$) the first resonance peak predicted
using approximate methods represents the rigorous numerical result
with a good relative accuracy (2\% of the central frequency).

In general VC integral equation method demonstrate better agreement
with the DDA calculations in comparison with the SI method (Fig.~\ref{fig:comparison_numeric}
and \ref{fig:map}). This can be best seen in figure~\ref{fig:map},
where the resonance frequencies of increasing order (from left to
right) against the wire length are depicted for nanowires with radius
$a=10nm$ (top panel) and $a=20nm$ (bottom panel). The VC integral
equation method (solid line), SI integral equation method (dashed
line) are presented. Dots represent DDA results. The brighter (darker)
curves are the resonances of even (odd) orders. All calculations are
done under slanting incidence ($\xi=\frac{\pi}{4}$). The better accuracy
of the VC integral equation method can be systematical traced in figure~\ref{fig:map}
especially for the structures with smaller aspect ratio.

\section{Conclusion}

Two alternative methods to solve the scattering problem on optical
nanowire antenna, the volume current integral equation (VC-IE) method
and the surface impedance integral equation (SI-IE) method are introduced.
In order to reduce the general 3D volume integral equation describing
the scattering problem to a simple semi-analytical 1D integro-differential
equation, both methods utilize solutions of the problem of plane wave
scattering on infinite cylinder. A regularization and discretization
scheme is proposed in order to transform integro-differential equations
into solely integral equation. This transformation enables to solve
the original problem without necessity to impose additional boundary
conditions at the nanowire edges. Numerical evaluation of the proposed
methods and their comparison with different numerically rigorous methods
is presented for scattering cross-section calculations. Gold nanowires
are analyzed at optical and near-infrared spectral range. The introduced
one-dimensional semi-analytical methods demonstrate good agreement
and superior numerical performance in comparison with rigorous numerical
methods.

\bibliographystyle{elsarticle-num}
\bibliography{biblio}

\end{document}